\documentclass{article}
\usepackage{spconf,amsmath,graphicx}

\usepackage{amsmath, bm}

\usepackage{xcolor}

\newcommand{\sstitle}[1]{\smallskip\noindent\textbf{#1.\/}}
\newcommand{\sititle}[1]{\smallskip\textit{#1\/}}

\newcommand{\revise}[1]{\textcolor{black}{#1}}

\newcommand{\eg}{e.\,g.,\ }
\newcommand{\ie}{i.\,e.,\ }

\usepackage{hyperref}
\usepackage{booktabs}
\usepackage{paralist}
\usepackage{float}

\usepackage{caption} 
\usepackage{subcaption}

\usepackage{fancyhdr}

\usepackage{amsmath}
\title{Prototype Learning for Interpretable Respiratory Sound Analysis}

\name{Zhao Ren, Thanh Tam Nguyen, Wolfgang Nejdl} %\thanks{XXX}
\address{L3S Research Center, Leibniz Universit\"at Hannover, Germany\\
{\small \tt \{zren, tamnguyen, nejdl\}@l3s.de}
\vspace{-6pt}
}
\begin{document}
%\ninept

\setlength{\belowdisplayskip}{0pt}
\setlength{\belowdisplayshortskip}{0pt}
\setlength{\abovedisplayskip}{0pt}
\setlength{\abovedisplayshortskip}{0pt}

\maketitle
% this sentence is for arxiv version only.
\thispagestyle{fancy}         
\fancyhead{}                     
\lhead{\scriptsize © 2022 IEEE. Personal use of this material is permitted. Permission from IEEE must be obtained for all other uses, in any current or future media, including reprinting/republishing this material for advertising or promotional purposes, creating new collective works, for resale or redistribution to servers or lists, or reuse of any copyrighted component of this work in other works.}       \chead{}
\rhead{Technical Report}
\lfoot{}
\cfoot{\thepage}  
\rfoot{}
\renewcommand{\headrulewidth}{0pt}  
\renewcommand{\footrulewidth}{0pt}

\begin{abstract}
Remote screening of respiratory diseases has been widely studied as a non-invasive and early instrument for diagnosis purposes, especially in the pandemic. The respiratory sound classification task has been realized with numerous deep neural network (DNN) models due to their superior performance. However, in the high-stake medical domain where decisions can have significant consequences, it is desirable to develop interpretable models; thus, providing understandable reasons for physicians and patients. To address the issue, we propose a prototype learning framework, that jointly generates exemplar samples for explanation and integrates these samples into a layer of DNNs. The experimental results indicate that our method outperforms the state-of-the-art approaches on the largest public respiratory sound database.

%, and is robust to class imbalance benchmarks.

%Remote screening of respiratory diseases has been widely studied as a non-invasive and early instrument for diagnosis purposes, especially in the pandemic. Such respiratory sound classification task has been realized with numerous deep neural network (DNN) models due to their superior performance. However, in high-stake domains where decisions can have significant consequences, it is desirable to develop interpretable models; thus, providing understandable reasons for physicians and patients. To address the issue, we propose a prototype learning framework, that jointly generates exemplar samples for explanation and integrates these samples into a layer of DNNs. 
%%By not separating the classification and explanation tasks, our framework achieves an interpretable respiratory sound classification that goes beyond the accuracy-interpretability trade-off. 
%%Unlike existing prototype-based explanation methods, our framework is robust to the multi-level domain of audio signals as well as the imbalance class problem.
%Our method outperforms the state-of-the-art approaches on the largest public respiratory sound database, and is robust to class imbalance benchmarks.
%%and the classification score is increased \todo{?$\times$ \%} compared to the baselines, 

%%Herein, this approach is verified on the International Conference onBiomedical Health Informatics 2017 challenge database.
\end{abstract}


\section{Introduction}
\label{sec:intro}

Respiratory sound classification is the task of automatically identifying adventitious sounds as a tool to assist physicians in screening lung diseases such as pneumonia and asthma~\cite{pramono2017automatic}. Unlike traditional auscultation, computer-aided auscultation of respiratory sounds provides a remote and non-invasive instrument for early diagnosis of patients at home or outside of hospitals. Owing to its promising prospect, respiratory sound classification has received considerable attention~\cite{tabatabaei2020methods,rocha2019open,song2021contrastive,chang2021transformer}.
%, and a number of models have been successively developed

Recently, deep neural networks (DNNs) have achieved great success in a wide range of areas. Due to their powerful capability, DNN-based models also have shown prominent performance in respiratory sound classification~\cite{song2021contrastive,yang2020adventitious}. However, a key limitation of these DNNs-based respiratory sound classification models is that they are not explainable by nature, especially in high-stake domains where decisions can have significant consequences like disease diagnosis. 
%For example, the pneumonia case study in \cite{caruana2015intelligible} showed that complex models overlooked how physicians explain predictive patterns in the data.

Prototype learning, emerging as a novel interpretable machine learning paradigm that imitates the human reasoning process, has attracted many recent works~\cite{chong2020towards,chen2019looks,li2018deep}. The basic idea of prototype learning is to explain the classification by comparing the inputs to a few \emph{prototypes}, which are similar examples in the application domain~\cite{rymarczyk2021protopshare,arik2020protoattend,ming2019interpretable}. Unlike posthoc explanation methods that only approximate original models~\cite{nauta2021neural}, prototype learning holds vast potential to improve the classification quality via nearest neighbor classifiers or kernel-based classifiers~\cite{zinemanas2021interpretable,chong2020towards}. With these efforts, the paradigm of prototype-based explanations has demonstrated some promising results showing that the so-called accuracy-interpretability trade-off~\cite{nauta2021neural} can be overcome.

However, despite the benefits of prototype learning, little attention is given for the audio domain. 
%In contrast to other classification tasks, sound classification, particularly respiratory sound analysis, is distinct because there are different levels of information encoded in audio data such as acoustic features and time-frequency representations. 
To address this issue and fully inherit the power of DNNs, we propose a prototype learning method for respiratory sound classification that integrates a prototypical layer into the training of an audio-driven convolutional neural network (CNN). Our framework takes input as the log Mel spectrogram of an audio signal as well as its delta and delta-delta because of their better performance than raw audio singals for DNN models~\cite{yan2021coughing}. Through a prototype layer that calculates the similarity between the internal feature map and the prototypes, the prototypes are learnt at the intermediate level to represent each class.
%\todo{Please summarized our framework here}.

%liu2020interpretable
Our work relates closely to the interpretable models such as k-nearest neighbors~\cite{coluccia2020robust}, attention mechanisms~\cite{arik2020protoattend,ren2020caanet}, and posthoc explanation methods~\cite{sudhakar2021ada,hase2019interpretable}. 
%These models improve the interpretability of a DNN by either show similar examples or the gradient activations of the DNN. 
However, it is difficult for humans to generalize from such interpretation due to the lack of quantifiable extent between the classification result and the explanation~\cite{chong2020towards}. 
%Our work also relates to posthoc explanation methods~\cite{sudhakar2021ada,hase2019interpretable}, which consider the original model as black-box and generate reverse-engineer information such as saliency maps and perturbations. However, most of these works are developed for image-based tasks. 
Our work relates most closely to Zinemanas et al.~\cite{zinemanas2021interpretable}, who proposed a network architecture that builds prototype-based explanation into an autoencoder. Unlike their model, our model employ cosine similarity between examples and prototypes, and apply attention-based similarity at the time-frequency level rather than frequency level only.
%... \todo{Please fill and edit this.} 
%Also, our model can overcome the imbalance class problem, which is common in disease diagnosis~\cite{tabatabaei2020methods}, by adjusting the number of prototypes per class.  
Also, our model can overcome the imbalance class problem, which is common in disease diagnosis~\cite{tabatabaei2020methods,chang2021covnet}.  

%The major contributions of this paper are summarized as follows. First, to the best of our knowledge, this is the first ever attempt to develop an interpretable respiratory sound classification framework. Second, we propose a prototype learning method to unify the power of DNNs as well as the explainability of case-based reasoning. Third, we conducted experiments on the largest public database for respiratory sound classification~\cite{rocha2019open}, and showed that our interpretable method outperforms non-interpretable counterparts as well as generates visually understandable explanations. Our model is also robust to the imbalance class problem.

To the best of our knowledge, this is the first attempt to develop an interpretable respiratory sound classification framework. We propose a prototype learning method to enhance the power of DNNs as well as the explainability of case-based reasoning. Constructing prototypes as explanations brings several benefits: (i) the learnt prototypes yield a concise representation and can be projected to the original data, (ii) it is easier to visually compare a classified audio sample and the exemplar examples (\ie prototypes), and (iii) a prototype can be a new case so that the physicians can understand more about the diseases.

%Paragraph 1: What is respiratory sound classification and why it is important.
%\edit{Respiratory sound classification means automatically identifying a recording, usually a respiratory cycle, as normal or containing adventitious sounds. It can overcome the subjective problem of conventional auscultation and has the potential for real-time monitoring.}
%
%Paragraph 2: Explain why we need explanation
%
%Paragraph 3: How we define the form of example-based explanation and what are the benefits
%
%Paragraph 4: Challenges in doing so and general idea of how we overcome these challenges

%\begin{figure}[!h]
%	\centering
%	\includegraphics[width=0.7\linewidth]{fig1.png}
%	\caption{Illustrative example for our approach}
%	\label{fig:example}
%\end{figure}

%\input{related}

\section{Methodology}
\label{sec:method}
\begin{figure}[!h]
    \centering
    \includegraphics[width=.95\linewidth]{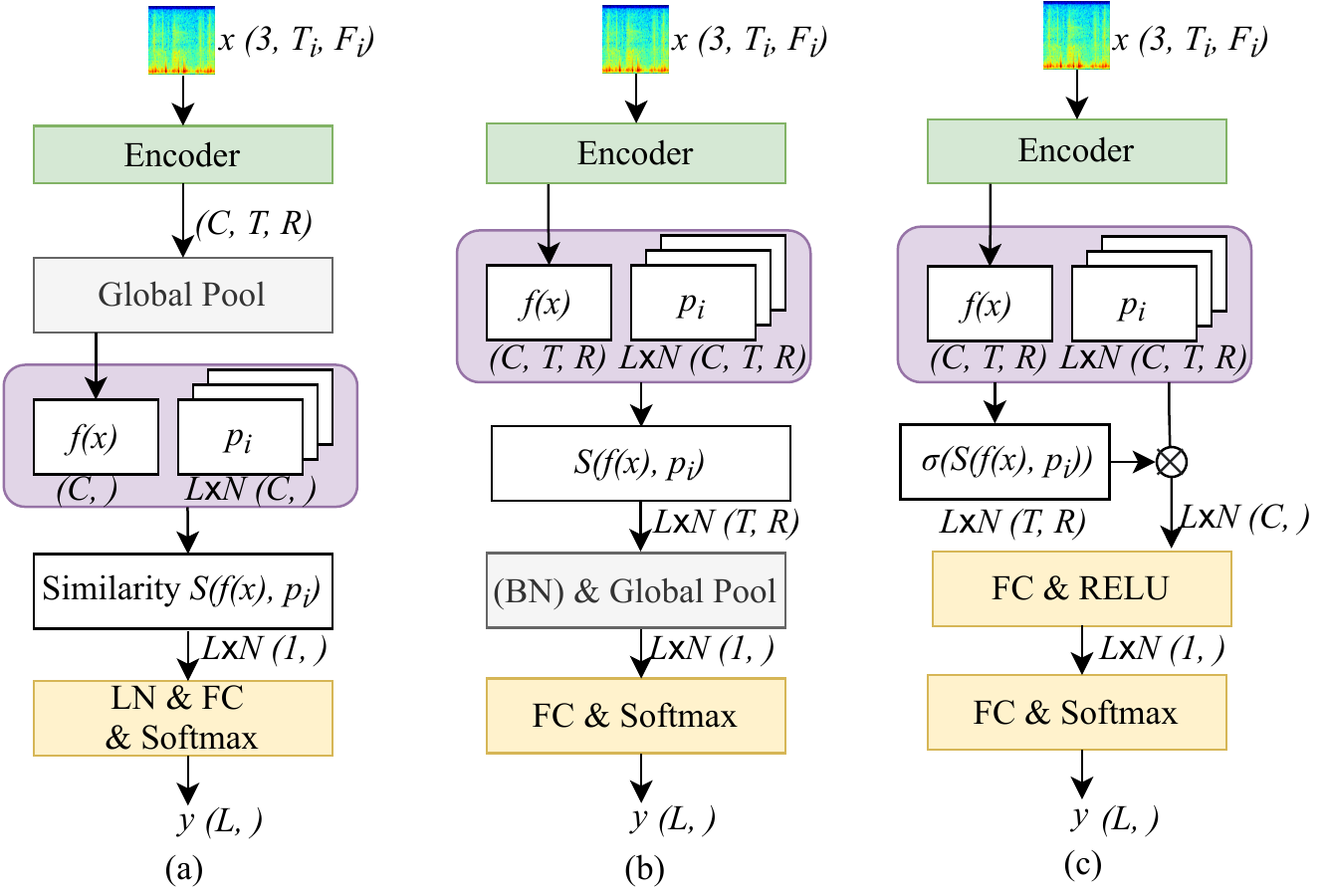}
    \vspace{-10pt}
    \caption{The frameworks of the prototype learning. (a) Prototype-1D. (b) Prototype-2D with vanilla similarity. (c) Prototype-2D with attention-based similarity. BN = batch normalisation, FC = fully connected layer, LN = layer normalisation, RELU = rectified linear unit. \revise{The input $x$ has the dimension of $(3, T_i, F_i)$, where $T_i$ is the number of time frames and $F_i$ is the Mel frequency bins. The other math symbols are corresponding to the definitions in this section.}}
    \label{fig:framework}
    \vspace{-10pt}
\end{figure}

With the extracted log Mel spectrograms as well as their deltas and delta-deltas as the input, three prototype learning approaches are employed in our work: i) Prototype-1D, ii) Prototype-2D with vanilla similarity, and iii) Prototype-2D with attention-based similarity (see \autoref{fig:framework}). Before learning the prototypes in each approach, a CNN model is employed as an encoder for analysing the respiratory sounds due to CNNs' strong capability of extracting highly abstract representations. 
%The CNN encoder is followed by a global max pooling and a fully connected (FC) layer for the classification task when a prototype layer is not trained. 
%Next, the prototype layers in the three above approaches are described in detail.

\subsection{Prototype-1D}
As the high-level representations include more class-related information than the low-level ones, the prototype layer is trained after a global max pooling layer for 1D prototypes (see \autoref{fig:framework}(a)). Given an instance $(x, y)$ ($x$: input, $y$: label), the intermediate representation before the prototype layer is denoted by $f(x)$. Through the prototype layer, a set of prototypes $P_l, l\in [1;L]$ are learnt, where $L$ is the classes' number. In each $P_l$, $p_i,\ i\in[1;N]$ is a prototype with the same size of $f(x)$, where $N$ is the number of prototypes for each class. The cosine metric is then used to measure the similarities between $f(x)$ and $p_i$:
\begin{equation}
    S_{1D}(f(x),p_i)=\frac{f(x)^Tp_i}{\parallel f(x)\parallel_2\parallel p_i\parallel_2}.
\end{equation}
The similarity is further fed into a layer normalisation (LN) layer, a fully connected (FC) layer, and a softmax layer for classification. Although the prototype-1D can generate prototypes, it is challenging for 1D prototypes to represent the time and frequency information.

\subsection{Prototype-2D}
As 2D prototypes can better represent the time-frequency information than 1D prototypes, the prototype layer is placed after the CNN encoder for the similarity measurements (see \autoref{fig:framework}(b-c)).  

\sstitle{Vanilla Similarity}
Similar to the prototype-1D learning approach, the calculated similarities are send to the next layers for the classification task (see \autoref{fig:framework}(b)).

\sititle{Element-wise Similarity}
aims to calculate the similarity scores between each pair of time-frequency bins in $(f(x))$ and $p_i$. When $f(x)$'s channel number is $C$ and its spatial size is $(T, R)$, the element-wise similarity is calculated by
\begin{equation}
    S_{2EV}(f(x), p_i)=\sum_{c=1}^{C}\frac{f^{c,t,r}(x)p^{c,t,r}_i}{\parallel f^{c,t,r}(x)\parallel_2\parallel p^{c,t,r}_i\parallel_2},
\end{equation}
where $t\in[1;T]$ and ${r\in[1;R]}$.

\sititle{Average Similarity} is to compute an average score across the similarities between all time-frequency bins of $p_i$ and each bin of $f(x)$. Since only part of an audio signal may contain the class-related characteristics, the average similarity is a score between each $f(x)$ bin and $p_i$:
\begin{equation}
    S_{2AV}(f(x), p_i)=\mathop{\mbox{avg}}\limits_{t'=1,r'=1}^{T, R}\sum_{c=1}^{C}\frac{f^{c,t,r}(x)p^{c,t',r'}_i}{\parallel f^{c,t,r}(x)\parallel_2\parallel p^{c,t',r'}_i\parallel_2}.
\end{equation}

\sititle{Maximum Similarity}
has the same idea of the average similarity, but select the most similar $p_i$ bin for each $f(x)$ bin for structuring the similarity scores at all time-frequency bins. The maximum similarity is computed by
\begin{equation}
    S_{2MV}(f(x), p_i)=\max_{t'=1,r'=1}^{T, R}\sum_{c=1}^{C}\frac{f^{c,t,r}(x)p^{c,t',r'}_i}{\parallel f^{c,t,r}(x)\parallel_2\parallel p^{c,t',r'}_i\parallel_2}.
    \label{eq:2mv}
\end{equation}

\sstitle{Attention-based Similarity}
Apart from the vanilla similarity, the attention-based similarity (see \autoref{fig:framework}(c)) is employed to learn weighted similarity scores. The calculated similarity scores are processed by a softmax function $\sigma(\cdot)$ for the attention feature maps. As the average similarity is computed as a global score across all $p_i$ bins for each $f(x)$ bin, it is not applicable for the attention-based similarity. Therefore, we introduce the element-wise and maximum similarities with the attention mechanism.

\sititle{Element-wise Similarity}
is defined by
\begin{equation}
    S_{2EA}=\sum_{t=1,r=1}^{T, R}\sigma(S_{2EV}(f(x), p_i))f^{c,t,r}(x)p^{c,t,r}_i.
\end{equation}
%where $\sigma(\cdot)$ is the softmax function.

\sititle{Maximum Similarity} is calculated at each $(f(x))$ bin $f^{c,t,r}(x)$ and its most similar $p_i$ bin $p^{c,t_{max},r_{max}}_i$ in Equation~(\ref{eq:2mv})
\begin{equation}
    S_{2MA}=\sum_{t=1,r=1}^{T, R}\sigma(S_{2MV}(f(x), p_i))f^{c,t,r}(x)p^{c,t_{max},r_{max}}_i.
\end{equation}

\subsection{Training}
During training the above prototype learning models, the prototypes are learnt as part of the model parameters. The loss function of the neural networks is finally defined by
\begin{equation}
    \mathcal{L}=\mathcal{L}_{NLL}+\alpha\mathcal{L}_{dv},
\end{equation}
\begin{equation}
\label{eq:diverseloss}
    \mathcal{L}_{dv}= \frac{\mathop{\mbox{avg}}_{l_1=1,l_2=1}^{L,L} S_{2AV}(P_{l_1}, P_{l_2})}{\mathop{\mbox{avg}}_{l=1}^{L} S_{2AV}(P_l, P_l)}, l_1\neq l_2,
\end{equation}
where $\mathcal{L}_{NLL}$ is the negative log likelihood (NLL) loss function when the neural networks' output in \autoref{fig:framework} is passed into a logarithm function, $\mathcal{L}_{dv}$ denotes the diverse loss function, and $\alpha$ is a constant value. $\mathcal{L}_{dv}$ aims to reduce the distances among prototypes which represent the same class and increase the distances among prototypes for different classes. \revise{The average similarity is experimentally used to evaluate the similarities between each pair of prototypes. In each set of prototypes $P_l$ for the class $l$, the similarities of each two different prototypes inside the $P_l$ are calculated and averaged, leading to the denominator in Eq.~(\ref{eq:diverseloss}); between each two different sets of prototypes $P_{l_1}$ and $P_{l_2}$, the averaged similarity is also computed as the molecular of Eq.~(\ref{eq:diverseloss}).}

\section{Empirical Evaluation}

\subsection{Experimental Settings}

\sstitle{Data}
The Scientific Challenge database released at the International Conference on Biomedical and Health Informatics (ICBHI) 2017~\cite{rocha2019open} is the largest publicly available collection of audio samples for respiratory sound classification. Totally $920$ audio recordings were collected from seven chest locations (\ie trachea, anterior left, anterior right, posterior left, posterior right, lateral left, and lateral right) of $126$ participants with four devices (\ie one microphone and three stethoscopes). The audio recordings have different sampling rates: $4$\, kHz, $10$\, kHz, and $44.1$\,kHz. All recordings derive $6\,898$ respiratory cycles, each of which was annotated with one of the four classes: \emph{normal}, \emph{crackle}, \emph{wheeze}, and \emph{both}, \ie \emph{crackle $+$ wheeze}. The database was split into a training set ($60\,\%$) and a test set ($40\,\%$) for the competition. 
To optimise the model hyperparameters, we further divide the training set into two subject-independent data sets: a train set ($70\,\%$) and a development set ($30\,\%$) (see \autoref{tab:datasets}).

%\todo{Please fill here}

%\autoref{tab:datasets}

\begin{table}[!h]
    \centering
    \scriptsize
    \vspace{-5pt}
    \caption{Distribution of the splitted train/development sets and the official test set in the ICBHI database. }
    \label{tab:datasets}    
    \vspace{-10pt}
    \begin{tabular}{l|p{1cm}p{1cm}p{1cm}p{1cm}p{1cm}}
    \toprule
         \#& \textbf{Normal} & \textbf{Crackle} & \textbf{Wheeze} & \textbf{Both} & $\bm{\sum}$ \\
         \hline
         \textbf{Train}& 1\,513 & \ \, 616 & 281 & 131 & 2\,541\\
         \textbf{Devel}& \ \, 550& \ \, 599 & 220 & 232 & 1\,601\\
         \textbf{Test} & 1\,579 & \ \, 649 & 385 & 143 & 2\,756 \\
         \hline
         $\bm{\sum}$ & 3\,642 & 1\,864 & 886 & 506 & 6\,898 \\
         \bottomrule
    \end{tabular}
    \vspace{-10pt}
\end{table}

\sstitle{Evaluation Metrics}
Although the database contains four labels, it is a common practice to differentiate abnormal cases (crackles, wheezes, and both) and normal cases. Therefore, the following standard benchmarks are used: \emph{sensitivity} (SE) -- equals to the number of true abnormal cases over the total number of abnormal cases, \emph{specificity} (SP) -- the ratio of true normal cases over normal cases, \emph{average score} (AS) -- is the official score of the ICBHI challenge~\cite{rocha2019open} and is the average of SE and SP. Due to class imbalance, we also report the \emph{unweighted average recall} (UAR) as the generic classification benchmark instead of \emph{accuracy}~\cite{song2021contrastive,ren2020generating}.

%\sstitle{Baselines} 
%To compare our proposed approach with the state-of-the-art (SOTA) methods on the ICBHI database, the implementation information of the SOTA methods are given as follows.
%\begin{compactitem}
%(1) \emph{MFCC-HG}~\cite{jakovljevic2017hidden}: The Mel-frequency cepstral coefficients (MFCCs) are extracted with their first derivatives as the input of the classifiers, which were structured with hidden Markov models (HMMs) and Gaussian mixture models (GMMs). 
%(2) \emph{MFCC-BDT}~\cite{chambres2018automatic}: The MFCCs from the audio signals are fed into a boosted decision tree (BDT) with four leaves for four classes.
%(3) \emph{STFT-WS}~\cite{serbes2017automated}: %Short-time Fourier transform (STFT) was applied to the channels produced by wavelet decomposition. 
%The frequency features integrated from the short-time Fourier transform (STFT) spectrograms and the statistical \& spectral features from the wavelet coefficients were classified by support vector machines (SVMs). 
%(4) \emph{STFT-WB}~\cite{ma2019lungbrn}: The STFT spectrograms and wavelet scalograms were extracted from the respiratory sounds and processed by a bi-ResNet model. 
%(5) \emph{STFT-RSESA}~\cite{yang2020adventitious}: The STFT spectrograms extracted from the audio signals were fed into a ResNet model with a squeeze-and-excitation (SE) block and a spatial attention (SA) block for the respiratory sound classification.
%\end{compactitem}

\sstitle{Implementation Details}
At the preprocessing stage, all audio recordings are resampled into $4$\,kHz due to the various sampling rates of the ICBHI database. A fifth butterworth bandpass filter ($100$\,Hz--$1\,800$\,Hz) is then applied to exclude noise components, \eg heart sounds, etc~\cite{rocha2019open}. The respiratory cycles with different durations are unified into audio signals with a fixed time length of $4$\,s. \revise{On the training stage, $4\,s$ segments are randomly selected from each data sample for improving the flexibility. On the testing stage, the middle $4\,s$ segments are selected to avoid the potential silence at the start and the end.} The log Mel spectrograms are further extracted from the audio signals with a window length of $256$, a hop length of $128$, and $128$ Mel bins, as they incorporate several properties of the human auditory system~\cite{martinez2014should}. %The static log Mel spectrograms, their deltas, and their delta-deltas are fed into the CNNs as the input. 

The CNN encoder is structured by four convolutional blocks with output channel numbers of $64$, $128$, $256$, and $512$, when each convolutional block consists of two covolutional layers with the same output channel number followed by a local max pooling layer with a kernel size of $2\times 2$. For the classification task, the CNN model with a CNN-encoder followed by a global max pooling layer and a FC layer is called `CNN-8'.

During training, the CNNs are optimised by an `Adam' optimiser with an initial learning rate of $0.001$ when the batch size is $16$. To stabilise the optmisation, the learning rate is reduced with a factor of $0.9$ at each $200$-th iterations. The training procedure is stopped at the $10\,000$-th iteration. To mitigate the class imbalance problem, each class in $\mathcal{L}_{NLL}$ is given a weight value which is inversely proportional to the samples' number of the class. The value of $\alpha$ is experimentally set to $0.1$.

\sstitle{Reproducibility Environment}
Our experiments are implemented at NVIDIA Geforce GTX 1080 Ti Graphics Cards. The PyTorch code is released at: \url{https://github.com/L3S/PrototypeSound}.

\subsection{End-to-end Comparison with SOTA Systems}

%To compare our proposed approach with the state-of-the-art (SOTA) methods on the ICBHI database, the implementation information of the SOTA methods are given as follows.
We compare our proposed approach with the following state-of-the-art (SOTA) methods on the ICBHI database.
%\begin{compactitem}
(1) \emph{MFCC-HG}~\cite{jakovljevic2017hidden}: The Mel-frequency cepstral coefficients (MFCCs) are extracted with their first derivatives as the input of the classifiers, which were structured with hidden Markov models and Gaussian mixture models. 
(2) \emph{MFCC-BDT}~\cite{chambres2018automatic}: The MFCCs are fed into a boosted decision tree with four leaves for four classes.
(3) \emph{STFT-WS}~\cite{serbes2017automated}: %Short-time Fourier transform (STFT) was applied to the channels produced by wavelet decomposition. 
The frequency features integrated from the short-time Fourier transform (STFT) spectrograms and the statistical \& spectral features from the wavelet coefficients were classified by support vector machines. 
(4) \emph{STFT-WB}~\cite{ma2019lungbrn}: The STFT spectrograms and wavelet scalograms were processed by a bi-ResNet model. 
(5) \emph{STFT-RSESA}~\cite{yang2020adventitious}: The STFT spectrograms were fed into a ResNet model with a squeeze-and-excitation block and a spatial attention block.
%\end{compactitem}
\autoref{tab:baselines} presents the result. Our approach performs better than all of the SOTA methods when comparing the AS scores. Our approach significantly outperforms the MFCC-HG approach ($p<.001$ in a one-tailed z-test).

\begin{table}[!h]
\centering
\footnotesize
\caption{Classification performance [\%] compared with the SOTA approaches on  the test set.}
\label{tab:baselines}
\vspace{-10pt}
\begin{tabular}{lllll} 
\toprule
  & SE & SP & AS & UAR  \\ 
\midrule
  MFCC-HG~\cite{jakovljevic2017hidden}  &--&-- & 39.56 &--     \\
MFCC-BDT~\cite{chambres2018automatic} & 20.81 & 78.05 & 49.43 &--      \\ 
STFT-WS~\cite{serbes2017automated}  &--&--&49.86 &--      \\
STFT-WB~\cite{ma2019lungbrn}  &31.12&69.20 &50.16 &--     \\
STFT-RSESA~\cite{yang2020adventitious}  &17.84&81.25&49.55 &--      \\
\hline
\textbf{Ours} & 27.78 & 72.96 & \textbf{50.37} & 36.16 \\
\bottomrule
\end{tabular}
\vspace{-10pt}
\end{table}

%\sstitle{Setting}
%Similar to section 2.5 of \cite{chen2019looks} and section V.A of \cite{chong2020towards}
%\sstitle{Results and discussion}
%\autoref{tab:baselines}
%Our prototype learning approach is compared to the above SOTA systems in \autoref{tab:baselines}. 

\subsection{Ablation Study}

\autoref{tab:ablation} shows our ablation study on the prototype layers and batch normalisation in the prototype-2D with the vanilla similarities. In general, the performance of most prototype learning variants is comparable to that of the basic CNN-8 model. Particularly, the UAR values are increased by several variants learnt by prototype learning, leading to higher SE values on the abnormal classes (\emph{crackle}, \emph{wheeze}, and \emph{both}). Both high accuracy and interpretability are reserved in our approach.

\begin{table}[!h]
    \centering
    \footnotesize
    \vspace{-3pt}
    \caption{Ablation study on the prototype layers and the batch normalisation (BN) when $N=1$. Pt = Prototype.}
    \label{tab:ablation}    
    \vspace{-10pt}
    \begin{tabular}{l|p{.5cm}|p{.5cm}p{.5cm}p{.5cm}p{.5cm}}
    \toprule
         \textbf{Performance [\%]} & \multicolumn{1}{c|}{\textbf{Devel}}  & \multicolumn{4}{c}{\textbf{Test}}    \\
         \midrule
         \textit{Variants} & AS  & SE & SP & AS & UAR \\ 
         \hline
         CNN-8  & 52.99   & 39.42 & 59.72 & 49.57 & 40.36 \\
         Pt-1D  & 51.44    & 46.73 & 45.85 & 46.29          & 43.34  \\
         \hline
         Pt-2D-EleSim-Van w/o BN  & 47.44  & 48.68 & 41.10 & 44.89 & 40.39 \\
         Pt-2D-EleSim-Van w/ BN  & 52.22 & 27.78 & 72.96 & \textbf{50.37} & 36.16 \\
         \hline
         Pt-2D-EleSim-Attention  & 48.36 & 46.22 & 37.62 & 41.92 & 40.06  \\
         \hline
         Pt-2D-AvgSim-Van w/o BN & 33.42  & 53.36 & 39.27 & 46.31 & 38.82  \\
         Pt-2D-AvgSim-Van w/ BN  & \textbf{55.43}  & 44.94 & 55.41 & 50.18  & 43.76  \\        
         \hline
         Pt-2D-MaxSim-Van w/o BN & 42.94 & 50.47 & 39.52 & 44.99 & 40.80 \\
         Pt-2D-MaxSim-Van w/ BN  & 11.85  & 61.94 & 00.00 & 30.97 & 35.36  \\
         \hline
         Pt-2D-MaxSim-Attention  & 50.81 & 49.96 & 37.62 & 43.79 & 43.50  \\ 
    \toprule
    \end{tabular}
    \vspace{-10pt}
\end{table}

The batch normalisation procedure is also analysed in our models with the vanilla similarities. In \autoref{tab:ablation}, batch normalisation leads to improvements for the vanilla element-wise similarity and the vanilla average similarity, which it results in very low performance (\ie $\mbox{SP}=0$) for the vanilla maximum similarity. 
%The reason might be that the maximum similarity could be calculated based on part of a prototype, leading to a challenge to optimise all time-frequency bins of the prototype. 
In this regard, we select the best batch normalisation setting for each vanilla similarity for further experiments.

\subsection{Sensitivity Analysis}

\begin{figure}[t]
    \centering
    \includegraphics[width=0.75\linewidth]{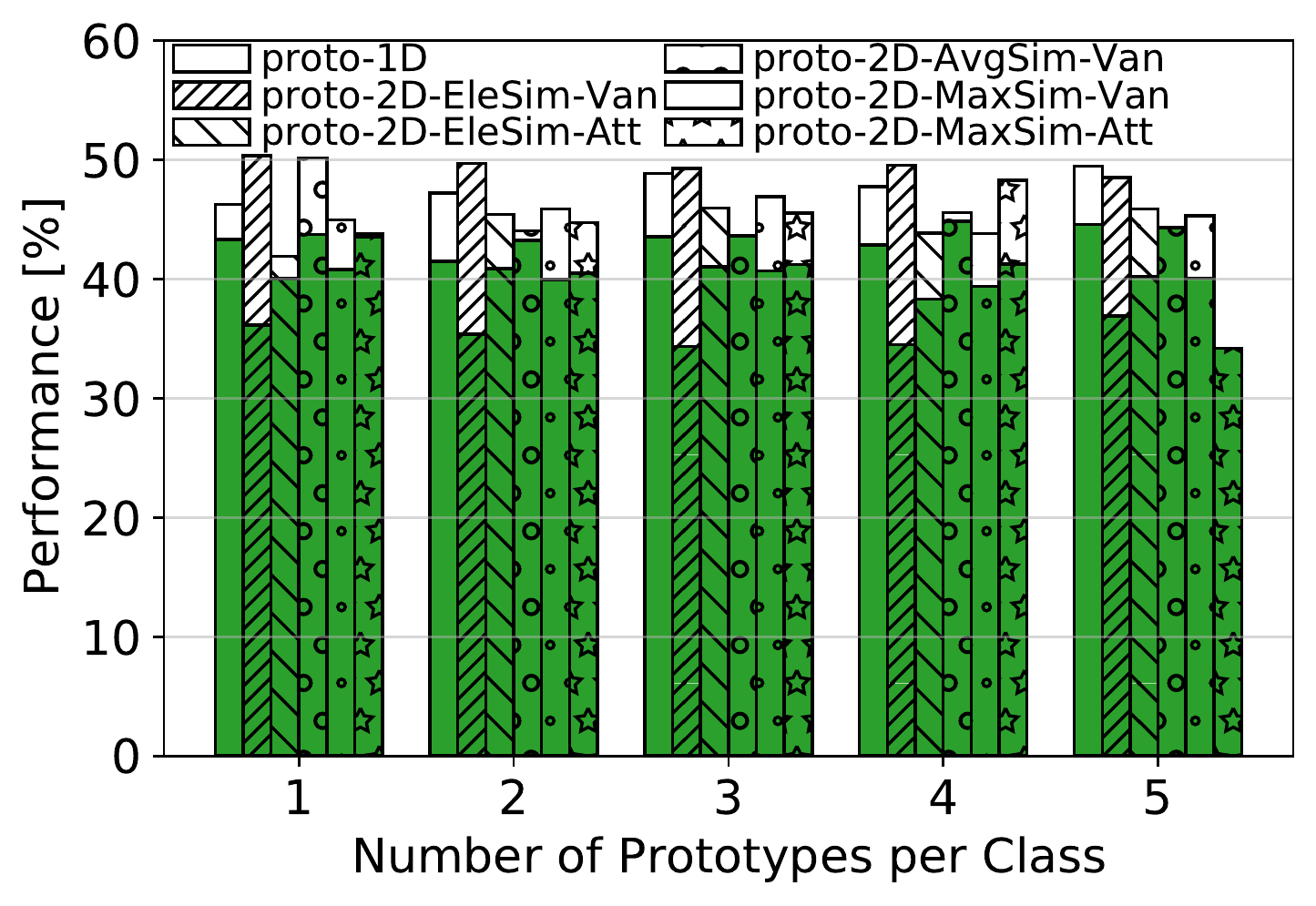}
    \vspace{-10pt}
    \caption{Performance [$\%$] comparison on the test set among different prototype learning approaches with a given number of prototypes per class. The green bars represent the UAR values, and the occluded white bars show the AS scores.}
    \label{fig:prototypes}
    \vspace{-10pt}
\end{figure}

\sstitle{Number of Prototypes}
We compare the effect of prototype numbers for each approach in \autoref{fig:prototypes}. The performances of the proposed models are comparable when $N$ varies from $1$ to $5$, indicating generating one prototype per class is sufficient.

\sstitle{Similarity Comparison} Prototype-2D with vanilla element-wise similarity mostly perform better than the other other approaches, perhaps due to its capability of generating prototypes with time-frequency information and less parameters than the models with the attention-based similarity. When comparing the three vanilla similarities, the vanilla element-wise similarity always outperforms the other two.

%\sstitle{Batch Normalization}

\subsection{Projection of Prototypes}
\begin{figure}[ht]
%\vspace{-10pt}
\begin{subfigure}{.12\textwidth}
  \centering
  \includegraphics[width=\linewidth]{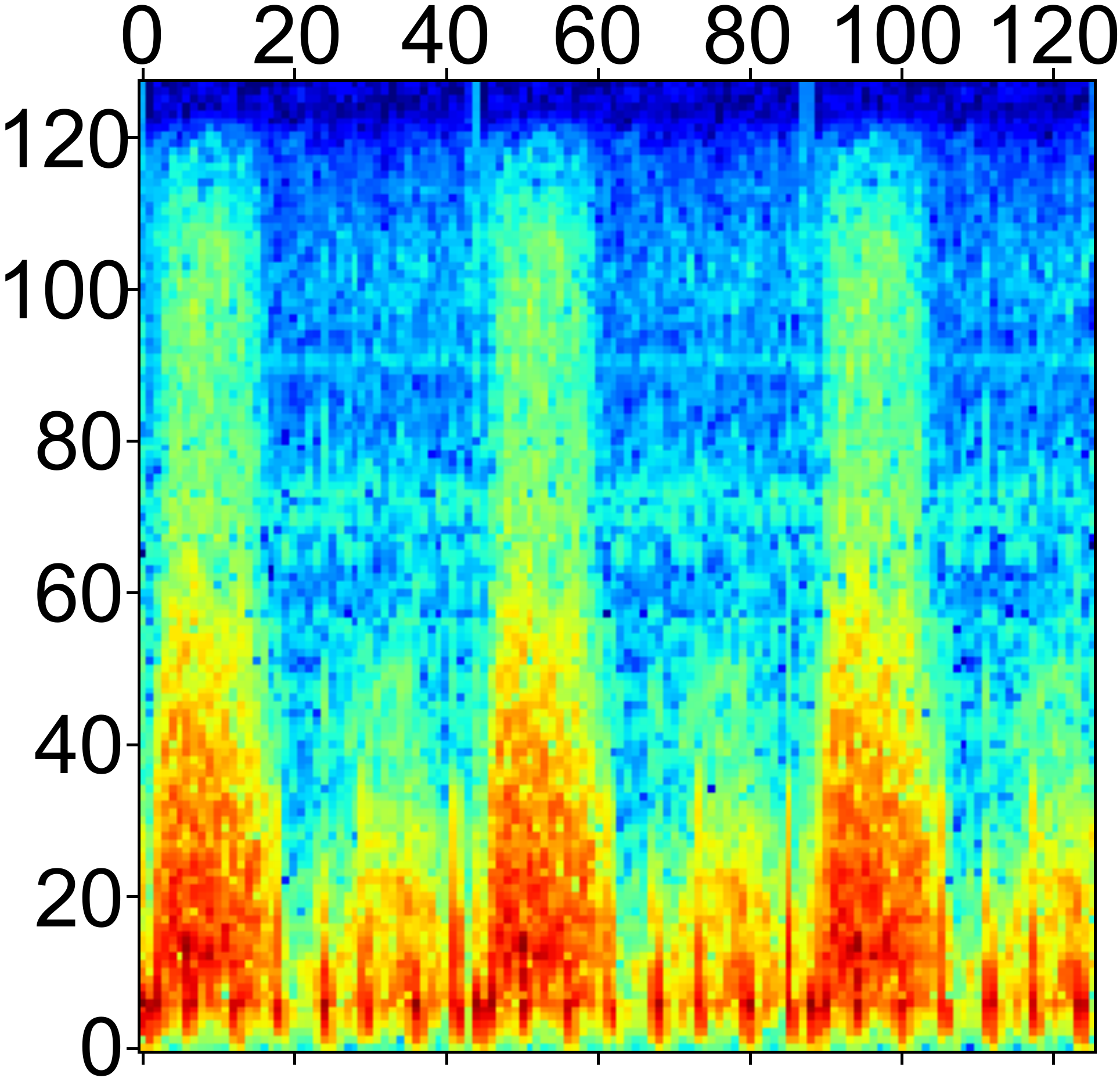}  
  \caption{Normal}
  \label{fig:sub-first}
\end{subfigure}
\hspace{-5pt}
\begin{subfigure}{.12\textwidth}
  \centering
  \includegraphics[width=\linewidth]{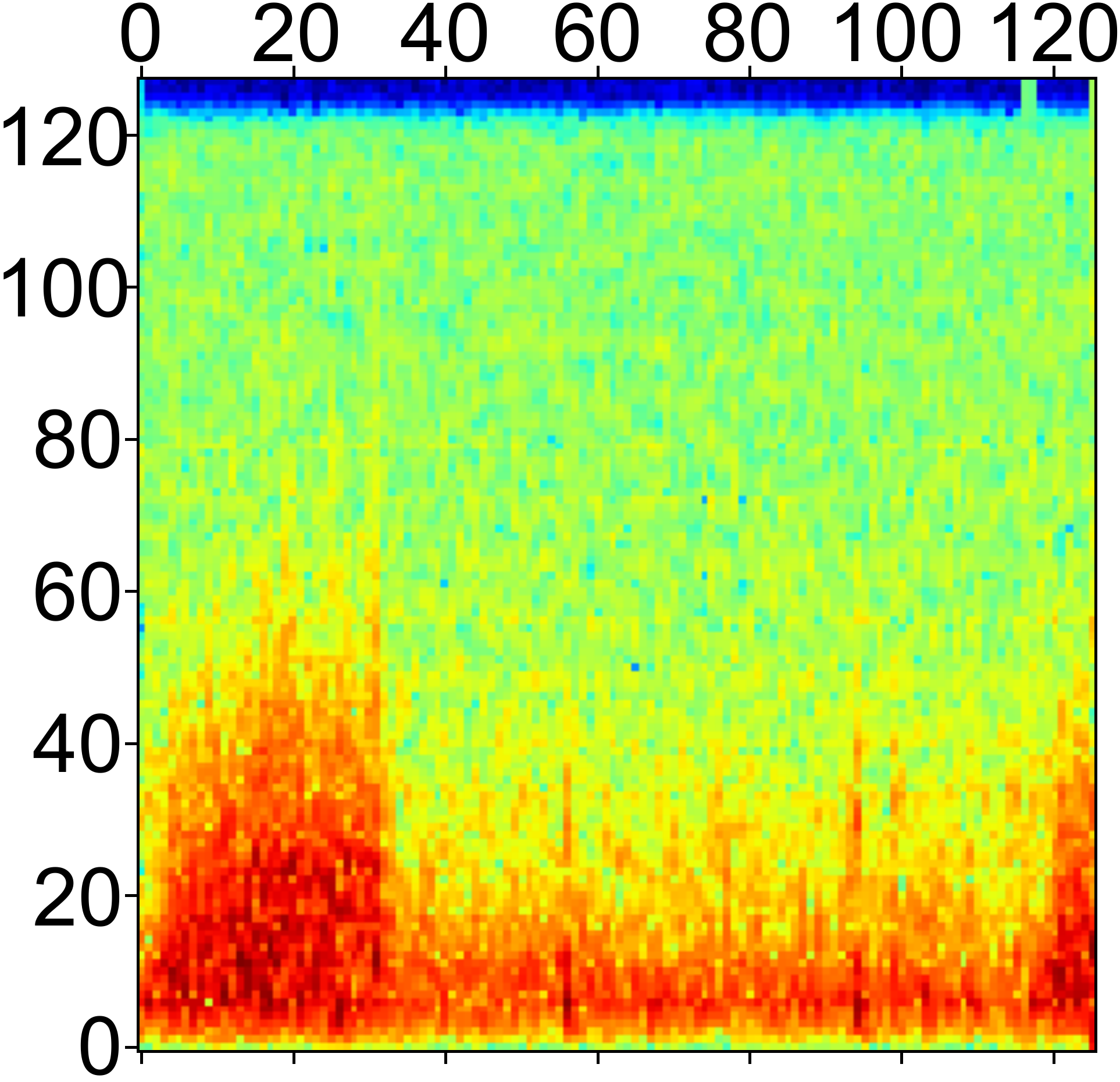}  
  \caption{Crackle}
  \label{fig:sub-second}
\end{subfigure}
\hspace{-5pt}
\begin{subfigure}{.12\textwidth}
  \centering
  \includegraphics[width=\linewidth]{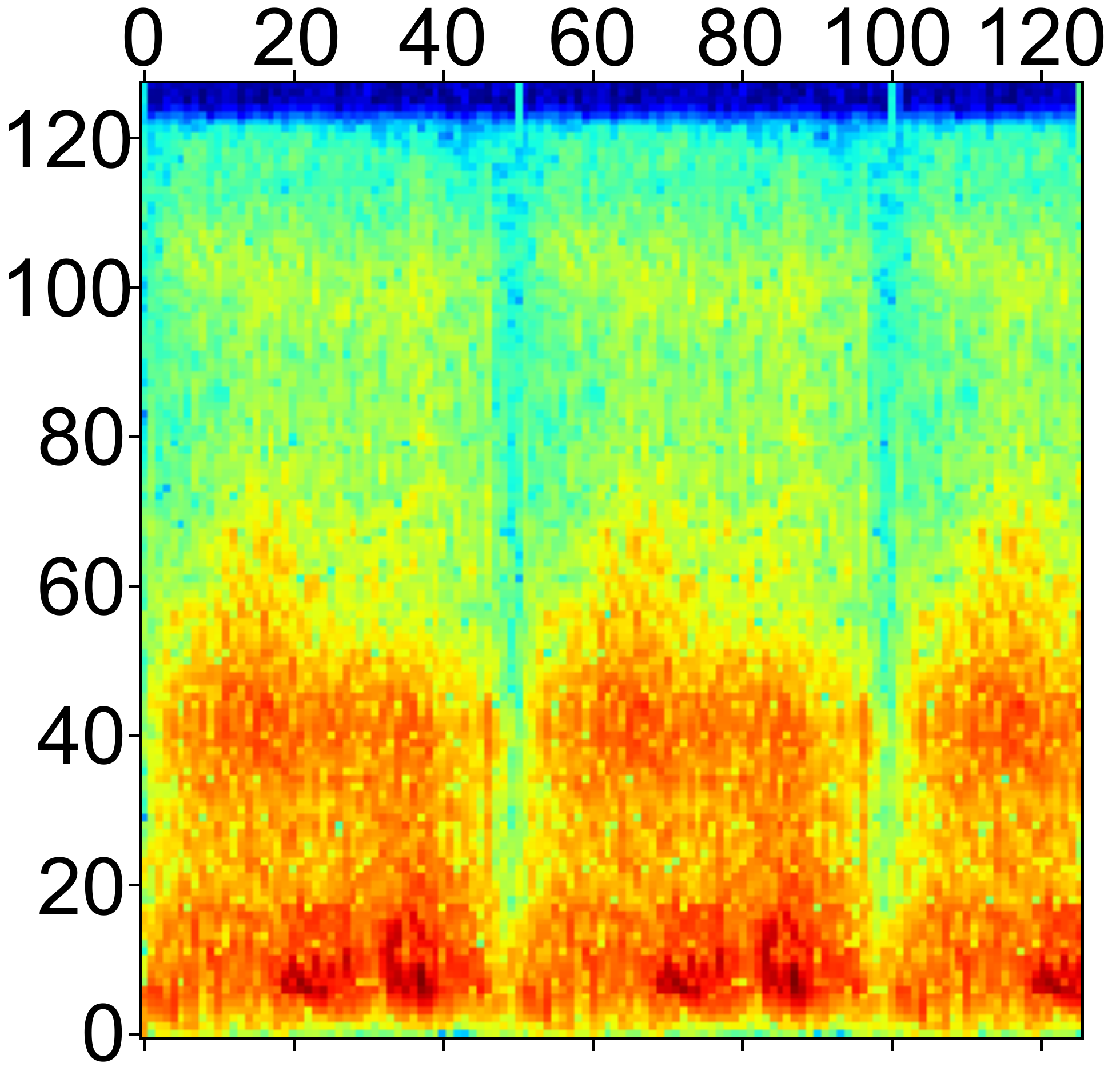}  
  \caption{Wheeze}
  \label{fig:sub-third}
\end{subfigure}
\hspace{-5pt}
\begin{subfigure}{.12\textwidth}
  \centering
  \includegraphics[width=\linewidth]{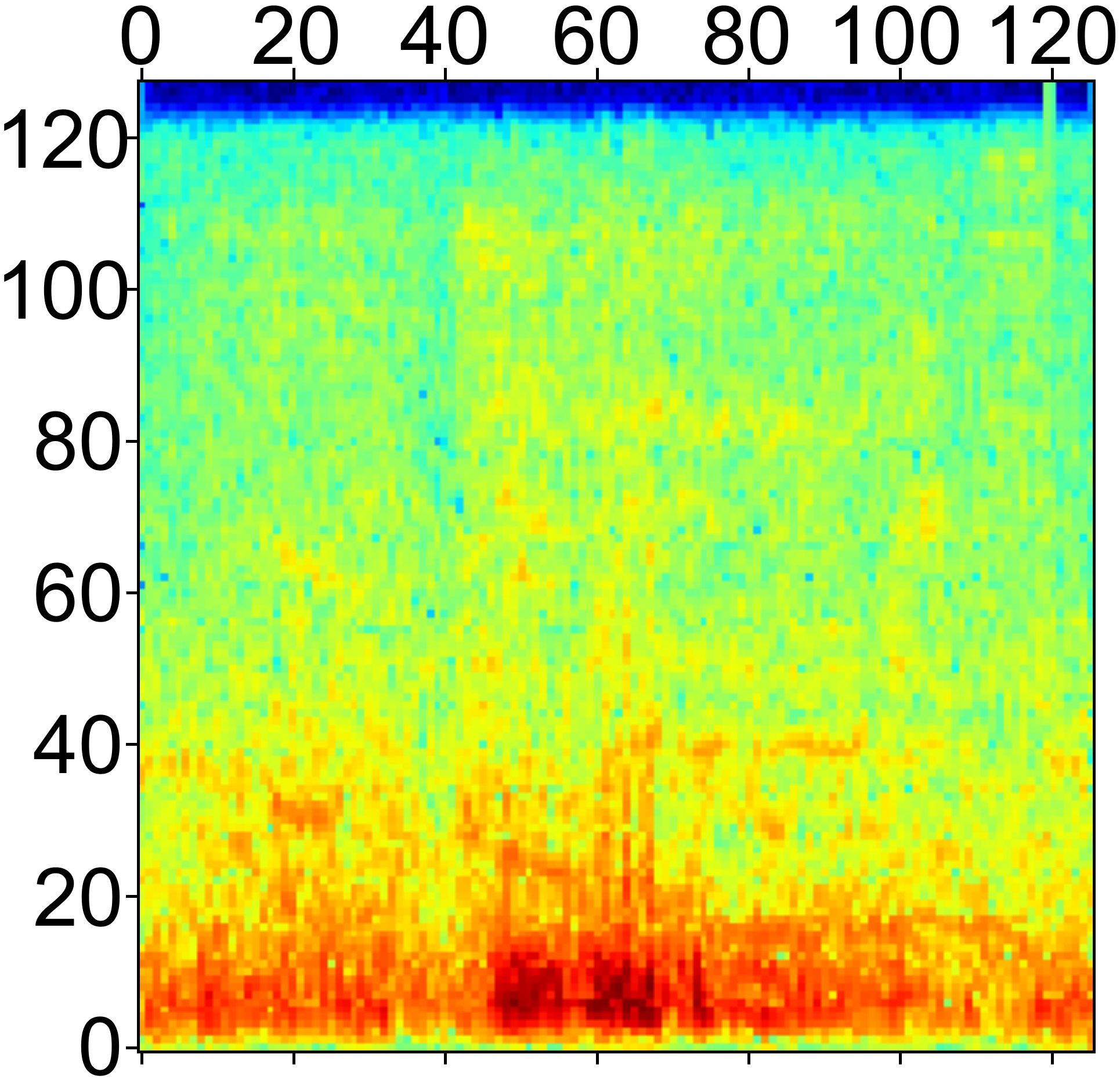}  
  \caption{Both}
  \label{fig:sub-fourth}
\end{subfigure}
\vspace{-10pt}
\caption{The closest log Mel spectrograms to the prototypes. The X-axis is time frame and the Y-axis is Mel frequency.}
\label{fig:projection}
\vspace{-5pt}
\end{figure}

As the generated prototypes contain multiple channels and have a small spatial size due to local pooling layers, it is challenging to visualize the prototypes. Hence, we project the prototypes to their closest inputs of the models \revise{by searching for the closest intermediate representation $f(x)$: 
\begin{equation}
\min dist_{j=1}^{J}(p_i, f(X_j)),\ dist(p_i, f(X_j))=e^{(-S)},
\end{equation}
where $X$ means all $J$ instances, $j$ is the index number. $dist$ is the distance, and $S$ denotes the similarity.
}
Herein, the log Mel spectrograms calculated by the projection procedure for our best model on the test set are depicted in \autoref{fig:projection}. The projection of prototypes is helpful to analyse the characteristics of each class of respiratory sounds. We can see that, the \emph{normal} respiratory cycles are regular, while the others are not. \revise{As crackle sounds are attributed to sudden bursts of air within bronchioles, therefore they are explosive and transient, and non-musical~\cite{serbes2011feature}. The example of \autoref{fig:projection}(b) can represent the above character of crackle sounds. Different from crackle sounds, wheeze sounds are continuous and commonly observed in patients with obstructive airways diseases, \eg asthma (AS) and chronic obstructive pulmonarydisease (COPD)~\cite{taplidou2007wheeze}. Musical wheeze sounds are sinusoidal sounds in time domain and are superimposed on normal breath sounds~\cite{taplidou2007wheeze}. In \autoref{fig:projection}(c), the regularity of the sounds can reflect the nature of wheeze sounds.} The \emph{wheeze} log Mel spectrogram has smaller coefficients on a range of Mel frequencies than the \emph{crackle} one, probably indicating the \emph{wheeze} sound is weaker. \revise{In \autoref{fig:projection}(d), the musicality of wheeze sound is difficult to be observed when wheeze and crackle sounds occur simultaneously.}

%\subsection{Qualitative Study}
%Any qualitative figures

%\subsection{Analysis of latent space}
%
%\sstitle{Setting}
%Show the quality of explanation, similar to section 2.6 of \cite{chen2019looks} and section V.B of \cite{chong2020towards}
%
%\sstitle{Results and discussion}

%\subsection{Reasoning process of our network}
%
%\sstitle{Setting}
%Similar to section 2.4 of \cite{chen2019looks}
%
%\sstitle{Results and discussion}

%\subsection{Some distinct experiment (Optional)}
%
%Some ideas: example ranking or example pruning (e.g. Figure 5(b) in \cite{chen2019looks})

\section{Conclusion}
\label{sec:conclusion}

Prototype learning paradigm, which is widely used for example-based explanation or case-based reasoning, recently has been transplanted to classification for jointly improving the classification performance and the result interpretability. 
%However, most prototype learning methods are under-investigated for respiratory sound classification, due to the multiple levels of semantic information in audio signals as well as the imbalance class problem in disease diagnosis. To address these issues, 
This paper developed a prototype learning framework for interpretable respiratory sound classification by generating prototypical feature maps that were integrated into the training of the predictive model. Not only increasing the predictivity, the learnt prototypes can introduce new cases to assist physicians in learning from automatic diagnosis and making informed decisions. In future work, we plan to explore other types of explanations such as concepts and criticisms~\cite{zinemanas2021interpretable} as well as reconstruct the original audio signals.

\sstitle{Acknowledgments}
This research was funded by the Federal Ministry of Education and Research (BMBF), Germany under the project LeibnizKILabor with grant No. 01DD20003.

%\vfill\pagebreak

\bibliographystyle{IEEEbib}
\bibliography{refs}

\appendix

\end{document}